# Addictive Facebook Use among University Students

Zeinab Zaremohzzabieh[1], Bahaman Abu Samah[1], Siti Zobidah Omar[1], Jusang Bolong[1] & Nurul Akhtar Kamarudin[1]

[1] Institute for Social Science Studies, Universiti Putra Malaysia, Malaysia

Correspondence: Zeinab Zaremohzzabieh, Institute for Social Science Studies, Universiti Putra Malaysia, Malaysia. Tel: 60-3-8947-1852. E-mail: zeinabzaremohzzabieh@gmail.com

**Abstract**

The Facebook has become an essential part of almost every university students' daily life, and while a large number of students seem to get benefits from use of the Facebook by exchanging information for educational goals, make friends, and other activities, the literature indicates that this social networking site can become addictive to some university students' users, which is one of the today's higher education matters. The aim of this study, therefore, is to explore the phenomenon of Facebook addiction among university students. Qualitative study using interview is used to gather data from nine International postgraduates of Universiti Putra Malaysia and the data established three themes (Compulsion to check Facebook, High frequency use, and Using Facebook to avoid offline responsibility) relied on the participants interviews. The findings from these three themes showed that these users considered their Facebook dependency, are known as salience, tolerance, and conflict. These results also lead to the conclusion that like most activities, moderation and controlled use are key. So, the best approach to preparing students for life in a knowledge-based society is to help them exercise self-control and achieve a level of balance when using Facebook. It is believed that the findings of this study would help other Facebook researchers by contributing to the limited academic literature in this area.

**Keywords:** addictive behavior, social networks, social network addiction, Facebook addiction, university students

## 1. Introduction

Social network sites like Facebook has become a global phenomenon and being one of the greatest importance means of communication. Today, more than 68.5% of young adults and teenagers use Facebook on a regular basis (Kuss & Griffiths, 2011). Facebook was developed in 2004 by Mark Zuckerberg, who was a Harvard University undergraduate at the time. Originally, membership was limited to Harvard students (2009) among whom the adoption rate was quite high (between 85 and 96%) (Lampe, Ellison, & Steinfield, 2006; Salaway, Caruso, & Nelson, 2008; Stutzman, 2009; Tufekci, 2008). However, by 2006, the platform was opened up to the world, and anyone aged 13 or older, with a valid email address, was allowed to join. Facebook is basically an online social network site in which users can share thoughts, ideas, pictures and other content with friends and family members, and to connect with either former or new friends, making the platform very popular with university students (Alexander, 2006; Boyd & Ellison, 2008; Ellison, Steinfield, & Lampe, 2007; Golder, Wilkinson, & Huberman, 2007; Joinson, 2008; Lampe et al., 2006; Luckin et al., 2009; Raacke & Bonds-Raacke, 2008; Salaway et al., 2008; Neil Selwyn, 2007; Stutzman, 2006). For example, Facebook is estimated to have more than 500 million members (Facebook, 2013), with the average user spending more than 20 minutes a day on the site (Cassidy, 2006; Needham & Company, 2007), and ranks as the most used site among university students (Lenhart, Purcell, Smith, & Zickuhr, 2010b; Violini, 2009)-to the degree that it would be difficult to find students who were not Facebook users. In a similar case, Shanaz (2010) found that the use of Facebook was increased particularly among female university students in Malaysian universities, and was the most popular site among the female students aged 18-24 years old. The other survey also found that 85 per cent of students were using Facebook to communicate with other students in their courses (Ophus & Abbitt, 2009). The findings of Stern and Taylor (2007) also show that 49 per cent of university students use Facebook, and that of those, 3 per cent spent more than two hours. Likewise, Sheldon (2008) found the greatest common uses of Facebook among 172 students with Facebook accounts (N = 160) being entertainment (M = 3.23, SD = 1.19), maintaining existing relationships (M = 3.64, SD = 1.24), and passing time (M = 3.88, SD = 1.23) (Peluchette & Karl, 2008; Raacke

& Bonds-Raacke, 2008; Salaway et al., 2008). However, despite the popularity of Facebook due to the great increase in use, speed, interactivity, and free Internet access, an amount of the student user population experiences some negative effects of excessive Facebook usage or are already captured in the 'web' of addiction. In its most common form, scholars have named this phenomenon "Facebook addiction" (Henrichs, 2009). It is similar to the habitual patterns of behavior associated with activities like gambling, shopping and Internet addiction. According to Stutzman (2005), users use Facebook to learn about each other and to develop social networks with their friends at university which are so vital for their socialization, this procedure can mark a turning point in an addictive behavior and it makes user to wasting time more and more on Facebook.

At this point it can be realized that university students remain a critical and unsafe position in terms of Facebook addiction. Because social networking sites take as an important part of their daily life and they provide a lot of helps to university students not only for a social purpose but also for educational purposes. Spending more and more time on Facebook unintentionally raise the chance of addiction in terms of Facebook use. It is required for students for the aspect of knowledge-based society to improve their capability in using technology for putting a plan into their work in a technological setting (Nalwa & Anand, 2003). Conversely, scholars concern about the consequences of alarming rate of mental and addiction and the problems related to heavy use of Facebook among university students. Recently, research that actually examined and measured specific behaviors related to addictive Internet use (Chou, 2001; Chuang, 2006; Mark Griffiths, 2000; Ng & Wiemer-Hastings, 2005; Quayle & Taylor, 2003; Tsai & Lin, 2003; Wan & Chiou, 2006; Yang & Tung, 2007; Young, Pistner, O'MARA, & Buchanan, 1999; Young, 1996, 1997). Nonetheless, we feel that this is a good time to fill the gaps in research around Facebook addiction, which has not yet a fully-fledged area of academic research. The majority of researchers on these topics have depended on the quantitative method and questionnaire technique to develop their ideas (Aghazamani, 2010; Shaffril, Samah, Uli, & D'Silva, 2011; Wang et al., 2011; Wilson, Fornasier, & White, 2010). For this purpose, this qualitative study is used to identify the extent of the problem of excessive Facebook usage among nine International postgraduates of Universiti Putra Malaysia. In addition to this, the study attempts to find out and discover what these Facebook addicts experience.

*1.1 Addictive Behavior and Social Network Sites*

Addictive behavior, referred to as impulse control disorders, is progressively known as remediable forms of addictions. It is also sometimes used for not substance-related addiction, soft or process addiction such as overeating, technology addiction, exercise, spiritual obsession, sexual addiction, compulsive shopping, and problem gambling, is completely different from chemical addiction (Albrecht, Kirschner, & Grüsser, 2007; Potenza, 2006; Shaffer & Hall, 1996). In these kinds of addiction, a compulsion happens many times by users to engage in some particular activity until it causes harmful consequences to their physical and mental health, or social life or wellbeing (Korolenko, 1992; Parashar & Varma, 2007; Stein, Hollander, & Rothbaum, 2009). It was classified for the first time by Tsezar Korolenko in Russia (Korolenko, 1992). He defines behavioral addiction as a disorder in which an individual has a tendency to escape from reality and can function by the ways of changing one's mental condition, is employed in a two basic ways:

1) Pharmacologically through the habit of psychoactive substances;

2) Non-pharmacological through the concentration on particular activities that are associated with subjectively pleasurable emotional conditions.

When analysing the addiction to Internet for example, a published study in 2011 by Iranian researchers (Alavi, Maracy, Jannatifard, & Eslami, 2011; Potenza, 2006) have shown that the similar molecular functions which make individuals into a behavioral not substance addiction are behind the compulsion to excessive Internet use, pushing Isfahan's University students into heightened level of psychological arousal, depression, low family relationships, and anxiety (Alavi et al., 2011). In the study, the core of addiction was not a substance but an alteration of the emotional cycles and its characteristics are as withdrawal/hedonic and tolerance symptoms, affective disorders, and problems in social relations.

Nowadays, with the increasing use of social network sites as the norm for how we communicate with each other and knowledge-sharing, psychologist and researches also come out with a new kind of social networking addiction as they believe that there is a correlation between how often one uses social network and a linkage to addiction (Ben-Ze'ev, 2004; Vallerand et al., 2003). According to Griffiths (2000b), social networking sites use might be a new form of soft addiction. Further, Vallerand et al. (2003) argued that the overuse of sites can become disruptive to daily life or lead to negative outcome such as loneliness, depression, anxiety, and phobias. This was most recently presented with a paper just published in 2011. In this paper, Kuss and Griffiths (2011) indicate that social network sites are crucial in terms of professional and academic opportunities, which clarify why some individuals are excessive social networking sites use. It can have series consequences for productivity of individual networks, they might wish to benefit. According to Deragon (2011), this kind of addictive behaviour is frequently seen to be a social state related to particular symptoms and signs. It may be caused by the internal factors (an absence of information and /or wisdom by people or groups in social networking sites use) or external factors (such as impact of social networking sites, simply using social networking sites in the incorrect way and for incorrect reasons) which is all about the production of meaningless things that do not improve the gradual persistence of individual network. With a variety of social network sites are available on the Web; students are attracted to neglect their study in preference for surfing these sites almost all day and hanging out online with friends and so many are already becoming addicted to the online rave of the moment-Facebook.

*1.2 Facebook Addiction*

Since Facebook is becoming a popular form of social networking sites, researchers have started to pay a great deal of attention to Facebook addiction problems among individuals. In recent years, Facebook users spend most of their time in the synchronous communication environment, engaging in interactive activities and some heavy users might be addicted. At this point, along with all the benefits Facebook brings, which may have unintended negative effects. Sometimes users cannot stop themselves from using Facebook to excess, allowing it to take up more and more of their time and thoughts, without any concrete outcome-like many addictions (Crandell et al., 2008; Lugtu, 2011). According to the American clinical psychologist, it may be reasonable to describe specifically of 'Facebook Addiction Disorder' (FAD) in detail because the criteria of addictive behavior such as neglect of work, mood modifying experiences, withdrawal, and tolerance emerge to be present in some individuals who excessive Facebook use. Meanwhile, Facebook addiction has similar signs to substance addictions. Users develop ignorance about personal life, an interest of using Facebook all the time, an experience for social escape; a number of defense mechanisms to hide addiction signs, a perceived loss of control, and a decline in pleasure over time (Kuss & Griffiths, 2011). For instance, psychologist and psychiatry in USA in May and June 2010 reported that from 1,605 participants surveyed on their social media habits, thirty-nine percent of them are self-described "Facebook addictions". Another sixty-one percent both admit to escape into the world of Facebook and browse what people are saying and doing in the middle of the night and prefer to see their friends on Facebook instead of face to face conversations (The Telegraph, 2012). There are a lot of similar cases like this study and almost all of them recognised that Facebook is so risky for turning into an addiction when it is employed unintentionally by people and there are so many side effect of these kinds of social networking sites can a come into picture more and more in a short time (Wilkinson, 2010). Excessive Facebook use has been found to harm psychological and social well-being of individuals and their personality (Harzadin, 2012). Larkshmi (cited in Alabi, 2013) claims that for those who are addicted to Facebook and their personal life truly interrupted by their uncontrolled activities on Facebook. As further clarified, the following is the checklist for Facebook addiction disorder (FAD):

1) Check your Facebook account in the morning;

2) Spend entire nights on the site, causing them to become tired the next day;

3) Spend hours a day on Facebook;

*4)* Whenever you are offline, you are just enjoying a daydream about the status updates and comment that have been received.

*1.3 Facebook Addiction among University Students*

In August 2007, comScore reported that Facebook is extremely valuable to university student in the United States due to its potential for academic purpose (Needham and Company, 2007). Florida University, For example, uses Facebook for academic purposes and teaching and learning, the University of Michigan uses Facebook to distribute news and connect its graduates (Boyd & Ellison, 2007; Cassidy, 2006; Needham and Company, 2007; Schwartz, 2009; Selwyn et al., 2008). These data portray Facebook as being an extremely significant among university students because it allows these students to improve cognitive, social competencies, a positive attitude towards learning and to develop communicate with the lecturer outside the classes (Kirschner & Karpinski, 2010; Pasek & Hargittai, 2009; Ross et al., 2009; Selwyn, 2009). Similarly, Yu, Tian, Vogel, & Chi-Wai Kwok (2010) noted that students generally thought the use of social networking sites such as Facebook is vital to obtain knowledge, social acceptance and support, which can lead students to higher levels of their self-esteem, performance skill, and satisfaction with university life. Chu and Meulemans (2008) realised also that the majority of graduates used Facebook to communicate with other students about their course, assignment, lecturer, and classes. In addition, Bosch (2009) shows that compared to university course sites, a large number of students are more engaged with Facebook to participate in discussions groups as Facebook has the potential to use as educational tool for work or group discussions. According to Stutzman (2008), North Carolina students preferred conducting discussions in Facebook, rather than the standard course management system.

On the other hand, some researchers have realised that there is significant positive association between problematic technology use and academic difficulties (Kubey, Lavin, & Barrows, 2001; Malaney, 2005). It has been proposed that social network sites use such as Facebook use may increase time spent online (Lenhart, Purcell, Smith, & Zickuhr, 2010a). Moreover, Facebook addiction and its effect on doing university work has been considered in numerous recent studies (Barrat, Hendrickson, Stephens, & Torres, 2005; Charnigo & Barnett-Ellis, 2013; Hafner, 2009; Martínez-Alemán & Lynk-Wartman, 2009; Sandvig, 2009; Stern & Taylor, 2007). The findings of Karpinski and Duberstein (2009) show that Facebook users (N = 148) from the Ohio State University (OSU) spent significantly less time studying (one to five hours versus 11 to 15 hrs., $p < .001$) than non-Facebook users, and had significantly lower grade point averages (GPA) ( 3.0-3.5 versus 3.5-4.0, $p < .001$). Karpinski & Duberstein also found that Facebook users studied less and got lower grades than non-Facebook users, while in a similar study, Boogart (2006) discovered a significant association ($p < .000$) among heavy Facebook use and lower grades.

Additional research is essential in order to determine whether college students' use of Facebook and/or other Internet sites is an issue of concern for higher education experts. Other researchers have found positive and/or not significant relationships between Facebook use and grades (Capano, Deris, & Desjardins, 2010; Hargittai & Hsieh, 2010; Pasek & Hargittai, 2009). Consistent with the findings, Kandell (Kandell, 1998) found that higher education students may be in danger of developing a Facebook addiction due to their developmental stage. He referred to Erikson (1963), who recognized forming identity and building close relationships as the primary developmental tasks of late adolescents and early adults. In another study, Sharifah et al. (2011) surveyed 380 female students in selected universities from Malaysia and found that there is a high relationship between motives of Facebook use (passing time, entertainment, and communication ) and Facebook addiction ($F=78.864$, $p<0.05$) at the 0.05 significant levels. Meanwhile multiple regression test displays passing time does contribute significantly towards Facebook addiction (0.24, $p <0.05$) and 17.3% variance in Facebook addiction is explained by passing time. The study concluded that these university female students are considered to be Facebook addicts. Most importantly, all this findings can encourage other scholars to further study the problems that could be the result of university students' excessive use of the Facebook. Facebook addiction among students is not a local phenomenon. University students all over the world pathologically may have any experience of social, academic, and psychological problem linked to online behavior.

## 2. Data Collection Methods

While earlier studies adopted the quantitative method through questionnaire technique and survey, the limitations of the methods were not appropriate for this study. Survey is a representative method to collects data from a large sample on accurate inquiries. Nevertheless, when addressing individual experience and self-concept, this kind of behaviour is rely on subconscious and requires in-depth exploration of the people' experiences rather than confirming or rejecting a preconceived hypothesis. According to Kozinets (2002), qualitative methods are mainly suitable for telling the rich detail and symbolic world which motivates needs, choice, and value. Therefore the topic of this research finds itself in qualitative phenomenological methodology because it provides the best opportunity to access the student experiences of Facebook addiction. In addition to the primary goal of this research, the participants for this study were nine international postgraduates from the Universiti Putra Malaysia (UPM) who reported using Facebook for 38 hours or more weekly for non-occupational and unnecessary proposes. Although the literature is still developing, the numbers of hours spent online has been employed to define Facebook addiction. Six male and three female students were selected based on the number of hours they self-reported spending on Facebook, and all participants were between the ages of 25 to 30. Their names and subjects are not indicated in order to protect their confidentiality. Finally, data was gathered from face to face semi-structured interviews to verify the conclusions in the research. This interview is useful when the objective is to recognize the constructs the interviewee uses as a foundation of their ideas and belief. The advantage of semi-structured interviews is that reduced the difficulty of item non-response and issue-based interviews of key participates offered full textured data (Easterby-Smith, Thorpe, Lowe, & McGarr, 1991). For semi-structured interviews, an interview protocol was notified about the rights of the interviewee that they could quit the interview at any time and also each interview time for each participant must be ranged from half an hour to an hour as it is proposed by Yin (1994). All of the interviews took place over two weeks.

*2.1 Data Analysis*

After the interview process was done, the researcher transcribed and analysed the findings related to the data base of this study. Three themes were identified as a result of the data analysis: compulsion to check Facebook, high frequency use, and using Facebook to avoid offline responsibility. Each of these findings is discussed in detail as follows:

*2.2 Theme One: Compulsion to Check Facebook*

Participants shared many statements linked to their opinions about Facebook use becoming compulsive and using Facebook more often than they primarily expected. All of them also reported the same statements that Facebook is the most important activity in their lives and starts to control their thinking. For example, one male participant stated, "I always try to open my mobile phone and go on Facebook and then I am thinking 'why I am doing this?'"

However, all male participants noted that they desired to be less Facebook usage, and neither of the female participants desiring to be less Facebook usage. One of the female participant stated, "I want to study more than past and get away from all of that." Of course, during interview, she took out her smartphone and attempted to found Wi-Fi to check her Facebook account. Other male participants told me "It becomes an addiction, I think I should stop this" in relation to desiring to avoid Facebook usage.

*2.3 Theme Two: High Frequency Use*

Many participants reported regularly using Facebook through the day and some of them reported they were using Facebook all the time. For example, one of the female participants had been spending approximately 5 hours per day checking her Facebook account. Other participants eventually spend all their free time on Facebook web page. One of the male participants stated "I used Facebook when I first wake up, throughout my workday, during lunch break". Female participants mentioned that many Facebook users post and share too much information on their page and it became an irresistible job for her to check the new status once a day. At that point she's overwhelmed with a lot of information and it is very difficult for her to catch up every read news or watch a news clip at once. Thus she spent lots of time out on Facebook, and devoted much of her day to catching up on any information, which missing out once she is back on Facebook.

*2.4 Theme Three: Using Facebook to Avoid Offline Responsibility*

Most of participants described choosing to visit Facebook as way of avoiding offline relationships, education/daily activities or responsibilities. The participants reported that excessive Facebook usage led them to many disorders in some daily activities and sleep, and finally led to change their life style compared to past means before overusing Facebook. One of the male participant stated that "I am tired and reluctant after I've started using Facebook." Other participant said "I ceased many of my routine activities, stayed a hostel most of the day in order to check my Facebook account". These participants always complain about the large amount of time they are spending on their Facebook page that effect their performance, they chat or check their message before something else that they need to do, and they try to stop using Facebook and fail. For example, one male participant recognised that the time he was spending on the site was threatening to his performance. He said, "I was at home trying to finish my assignment when I realised my habit to use Facebook have getting out of my hand" and then continued by saying that "I couldn't resist checking my Facebook and before I know it, few minutes had turned into many hours and I had not done my assignment at the end of a day". The researcher recognised that he is among the few to have accepted excessive Facebook usage as problematic. Similarly, other female participant noted that "I waste my time there and I neglect doing something that I should be doing some work". "Even though I like using Facebook, I don't get to do it all the time with university, and that's the just thing that lead me to be socialized and also it's not an important experience as what my offline relationship." she said.

## 3. Discussion and Recommendation

This study explored the phenomenon of Facebook addiction among nine International postgraduates. The data established three themes relied on participants interviews. The findings from these three themes (Compulsion to check Facebook, High frequency use, and Using Facebook to avoid offline responsibility) showed that these users considered their Facebook dependency, are known as salience, tolerance, and conflict (Brown, 1993; Griffiths, 1996, 2005). Based on the first themes, the students use Facebook compulsively and frequently struggle when they attempt to solve their problem and use Facebook less. This sign of Facebook addiction are often referred to as a "salience" and it can occurs when a specific activity such as social networking site becomes part of person's daily routine life and begins to control that person's feelings, and behaviour (Griffiths, 2005). In addition, the researcher have examined whether the Facebook usage can lead to other harmful effects or become disruptive to everyday life (Vallerand et al., 2003). Griffiths (2000a) claimed that this sign of the social networking sites dependency might be a new form of addiction or simply "tolerance". The tolerance refers to the fact that increasingly larger amounts of time are spent on social networking site to achieve the original effects (Griffiths, 2005). Young (1996) confirmed that computer addictions desire to increasing amounts of computer usage in order to feel happy with the experience. Finally, overusing Facebook allows the students to ignore offline responsibilities, relationships, and other activities. In line with this, social networking site addictions reported moderate to severe distress in real-life communities, academic performance, and work which are stated as a "conflict" (Griffiths, 2005; Kuss & Griffiths, 2011). According to Griffiths (2005) and Brown (1993), the conflict is the activity causes conflict in academic settings, interpersonal relationships, and other activities. These results also lead to the conclusion that, despite Facebook's potentially destructive effects, Facebook is, as Facebook itself proclaims, still a great way for students to "keep up with friends, upload an unlimited number of photos, share links and videos, and learn more about the people they meet" (Facebook, 2010). However, like most activities, moderation and controlled use are key, and the best aspect of preparing students for a modern world, therefore, is helping them learn self-control and balance when using technology.

In addition to the result of the study, it is essential to therapists treating clients and researchers in the field of technology addiction and counselling. One of the suggested solutions for understanding the phenomenon and prepare therapists for the future is to develop programs for awareness of the disorder by a two-day workshop and a lecture. Through workshop, consular admitting Facebook addiction as a problem and increasing awareness about the risk factors of Facebook misuse at university. Another solution for avoiding neglect of academic, work, and demotic responsibilities have all been identified as consequences of heavy Facebook usage is to provide counselling service through workshop for students who are either addicted or not to Facebook at campus. A key answer to protect overuse Facebook is to abstain from Facebook for one day of sport activity which helps users to reduce their need to find themselves on the Facebook, to have new experience in real life, and to make new friends off-line at campus-besides;

1) University counsellors can provide small group or individual counselling for students with Facebook addiction.

2) The emergence of a new social-psychology journal is being developed that will focus upon aspect of Facebook use and addiction/disorder.

3) Such effective recovery programmes, continued research is essential to better understand the underlying motivation of Facebook addiction.

4) Future research should develop treatment protocols and conduct outcome studies for effective management of these symptoms

These points, therefore, can use as a perfect solution to this entire problem and leads the students to be more productive and healthy while they are on the Facebook, to have a heightened appreciation of interface to real and virtual worlds, to be self-control when they are on Facebook, and to push them forward to develop their responsibility in relation to society and make awareness of Facebook addiction.